**Title**

Tensor-valued and frequency-dependent diffusion MRI and magnetization transfer saturation MRI evolution during adult mouse brain maturation

**Authors & Affiliations**

*Naila Rahman[1,2], *Jake Hamilton[1,2], Kathy Xu[3,4], Arthur Brown[3,4], Corey A. Baron[1,2]

1. Centre for Functional and Metabolic Mapping (CFMM), Robarts Research Institute, University of Western Ontario, London, Ontario, Canada

2. Department of Medical Biophysics, Schulich School of Medicine and Dentistry, University of Western Ontario, London, Ontario, Canada

3. Translational Neuroscience Group, Robarts Research Institute, Schulich School of Medicine and Dentistry, University of Western Ontario, London, Ontario, Canada

4. Department of Anatomy and Cell Biology, University of Western Ontario, London, Ontario, Canada

*Naila Rahman and Jake Hamilton contributed equally as co-first authors.

**Corresponding author:** Naila Rahman (nrahman25@uwo.ca)

**Abstract**

Although rodent models are a predominant study model in neuroscience research, research investigating healthy rodent brain maturation remains limited. This motivates further study of normal brain maturation in rodents to exclude confounds of developmental changes from interpretations of disease mechanisms. 11 C57Bl/6 mice (6 males) were scanned longitudinally at 3, 4, 5, and 8 months of age using frequency-dependent and tensor-valued diffusion MRI (dMRI), and Magnetization Transfer saturation (MTsat) MRI.

Total kurtosis showed significant increases over time in all regions, which was driven by increases in isotropic kurtosis while anisotropic kurtosis remained stable. Increases in total and isotropic kurtosis with age were matched with increases in MTsat. Quadratic fits revealed that most metrics show a maximum/minimum around 5-6 months of age. Most dMRI metrics revealed significantly different trajectories between males and females, while the MT metrics did not. Linear fits between kurtosis and MT metrics highlighted that changes in total kurtosis found throughout normal brain development are driven by isotropic kurtosis, while differences in total kurtosis between brain regions are driven by anisotropic kurtosis.

Overall, the trends observed in conventional dMRI and MT metrics are comparable to previous studies on normal brain development, while the trajectories of our more advanced dMRI metrics provide novel insight. Based on the developmental trajectories of tensor-valued dMRI and MT metrics, our results suggest myelination during brain maturation is not a main contributor to microscopic diffusion anisotropy and anisotropic kurtosis in axons. For studies that only calculate total kurtosis, we suggest caution in attributing neurobiological changes to changes in total kurtosis as we show here constant anisotropic kurtosis in the presence of increasing myelin content.

**Introduction**

Rodent models are a predominant study model in basic neuroscience research (1), with applications in ageing and Alzheimer's disease (2), traumatic brain injury (3), brain tumours (4), and other neuroscience research avenues. Most studies consider adulthood in rodents at 2-3 months of age, assuming a steady state condition of adulthood (5). However, Hammelrath et al. (5) and Mengler et al. (6) demonstrated that myelination continues to increase past 3 months of age in rodents, using T2-weighted MRI, diffusion MRI (dMRI), and histology. Many MRI studies investigated early postnatal neurodevelopment in rodents (7–16), but few have explored normal brain maturation after 3 months of age. This motivates further study of normal brain maturation in rodents to exclude confounds of cerebral developmental changes from interpretations of disease and injury mechanisms.

As neurobiological changes are challenging to track longitudinally using histology, dMRI provides a non-invasive means to capture changes in brain microstructure during development, aging, disease, and injury by probing the diffusion of water molecules (17). The most widely used dMRI technique is diffusion tensor imaging (DTI), which assumes the dMRI signal is entirely characterized by Gaussian diffusion (18) and utilizes a diffusion tensor model to estimate metrics including mean, axial, and radial diffusivity (MD, AD, and RD), and fractional anisotropy (FA). Diffusion kurtosis imaging (DKI) provides more information about the underlying tissue via the diffusion kurtosis, which quantifies the deviation from Gaussian diffusion (19). However, both DTI and DKI are unable to distinguish between microstructural changes and neuron fiber orientation dispersion (18,20), reducing their specificity to microstructural changes in brain regions with crossing fibers.

To reduce orientation dispersion effects on diffusion measurements, tensor-valued diffusion encoding (21–23), which varies the shape of the b-tensor (describes the strength of diffusion weighting along each direction) to vary the sensitivity to diffusion anisotropy, was developed. The utility of tensor-valued encoding stems from the fact that diffusion kurtosis characterizes the heterogeneity of sub-voxel sources of diffusion coefficient heterogeneity (19). Linear tensor encoding (LTE), which is the conventional method of encoding diffusion along a single diffusion direction at a time, is sensitive to sources of diffusion coefficient heterogeneity from both isotropic and anisotropic microstructural components, while spherical tensor encoding (STE), which encodes diffusion equally along all directions at the same time, is only sensitive to isotropic sources of sub-voxel diffusion heterogeneity. Accordingly, tensor-valued dMRI distinguishes between different sources of kurtosis and allows for computation of microscopic fractional anisotropy (µFA), which reports water diffusion anisotropy independent of the neuron fiber orientation dispersion (20,24,25). Previous studies have shown that tensor-valued dMRI provides better sensitivity than conventional DTI in distinguishing between different types of brain tumours (20), assessment of multiple sclerosis lesions (26,27), and detecting white matter microstructure changes associated with HIV infection (28). The normalized signal intensity of powder-averaged (i.e., average over all diffusion directions) dMRI acquisitions of a multiple Gaussian component system can be represented by the cumulant expansion (19,25):

$$\ln\left(\frac{S}{S_o}\right) = -bD + \frac{1}{6}bD^2K \ldots \quad (1)$$

where S is the powder-averaged signal, $S_o$ is the mean signal with no diffusion encoding, b is the b-value, D is the diffusivity, and K is the kurtosis of the powder-averaged signal. By fitting Eq. (1) with the powder-averaged LTE and STE signals ($S_{LTE}$ and $S_{STE}$ respectively), the total kurtosis ($K_{total}$), which is the conventionally reported mean kurtosis measure in DKI, and isotropic kurtosis ($K_{iso}$), can be calculated, respectively. $K_{iso}$ is a measure of the variance in the magnitude of diffusion tensors (i.e., MD), which can be related to cell size heterogeneity (20). Additionally, by subtracting $K_{iso}$ from $K_{total}$, the kurtosis arising from diffusion anisotropy, $K_{aniso}$, can be calculated:

$$K_{total} = K_{iso} + K_{aniso} \quad (2)$$

Subsequently, µFA can be expressed in terms of $K_{aniso}$ by (25):

$$\mu FA = \sqrt{\frac{5}{4}}\sqrt{K_{aniso}} \quad (3)$$

Conventional dMRI sequences are also limited to probing length scales on the order of 10-30 μm due to hardware constraints (29), as they rely on the pulsed gradient spin echo (PGSE) sequence to encode diffusion. To probe smaller length scales, the oscillating gradient spin echo (OGSE) sequence was developed to modify sensitivity to cellular length scales (30). OGSE dMRI has helped to identify neurite beading as a mechanism for contrast after ischemic stroke (30,31), and has shown increased sensitivity compared to conventional PGSE dMRI in the assessment of hypoxia-ischemia (33) in rodents, and in various pathologies in humans (34–36). By varying the frequency of the gradient waveform (i.e., frequency-dependent dMRI), which is inversely related to diffusion time, OGSE encoding allows different microstructural length scales to be probed. For increasing diffusion times (lower oscillating gradient frequencies), the molecules travel greater distances and interact with more barriers such as cell membranes, resulting in lower observed diffusivity values and higher FA values (37). Subsequently the difference in these DTI metrics between the highest and lowest frequencies applied (ΔMD, ΔAD, ΔRD, and ΔFA) can be used to examine their dispersion with frequency and provide greater insight into tissue microstructure (38,39). Additionally, evidence of a linear dependence of MD on the square root of frequency has been demonstrated in healthy and globally ischemic rodent brain tissue (40) and healthy human white matter (41,42). Therefore, the diffusion dispersion rate (Λ) can be calculated using a power-law model as:

$$\mathrm{MD_f} = \mathrm{MD_0} + \Lambda f^{0.5} \qquad (4)$$

where $\mathrm{MD_f}$ is the MD at OGSE frequency, f, and $\mathrm{MD_0}$ is the MD at f = 0, which is the conventional PGSE sequence (38,39,41).

As myelination is an adaptive process that continues throughout adulthood, changes in myelin content throughout healthy brain maturation are important to consider when comparing to effects of disease/injury. As myelin is MR-invisible in diffusion-weighted scans, recent studies have applied both dMRI and myelin-sensitive methods for a more comprehensive view of microstructural changes (43–45). Sensitivity to myelin can be encoded in the MR signal by probing the magnetization transfer (MT) effect of bound water molecules by quantifying the MT ratio (MTR) (46). However, MTR has been shown to be sensitive to sequence parameters, flip angle inhomogeneities, and T1 effects (47,48). MT saturation (MTsat) imaging has been shown to reduce these sensitivities and increase specificity to changes in myelin content (47). Additionally, changes in DTI, DKI, and µFA metrics have been correlated with changes in myelin content and/or integrity of the myelin sheath (49–51), however, the biophysical nature of how myelination impacts these metrics remains unclear. Many studies have correlated an increase in total kurtosis with increased myelin content, however, whether this is the anisotropic or isotropic diffusion component ($K_{aniso}$ and $K_{iso}$, respectively) remains to be determined.

In humans, a number of MRI studies have been conducted on brain maturation and aging across the entire lifespan, most involving large multi-center datasets (52–64). The studies have included structural MRI, DTI, DKI, NODDI (a biophysical dMRI model), and myelin-sensitive MR techniques, with quantitative and volumetric analyses. Lifespan patterns of quantitative and volumetric analyses were widely reported to follow U-shaped or inverted U-shaped trajectories, including brain regional variations in these trajectories. For example, MD demonstrated a U-shape, while FA demonstrated an inverted U-shape, showing that brain maturation continues until middle age followed by a phase of degeneration at older ages. Most studies have reported on MD and FA, while only Latt et al. (55) and Das et al. (56) have reported on diffusion kurtosis changes over the human lifespan, and Cheung et al. (65) and Han et al. (66) have reported on diffusion kurtosis evolution in rats, up to 3 months and 13 months of age, respectively. $K_{total}$ has been shown to follow the inverted U-shape trajectory through normal development in humans and rodents. However, the sources of kurtosis that drive these changes remains unclear, so it is necessary to disentangle how each kurtosis component is affected by neurobiological changes during ageing to examine what changes in $K_{total}$ could indicate. Importantly, frequency-dependent and tensor-valued dMRI metrics, which provide improved sensitivity and specificity to microstructural changes, have not been explored in terms of normal brain maturation.

As microstructural MRI studies in rodents are becoming more prevalent in neuroscience research, our aim was to investigate how frequency-dependent and tensor-valued dMRI, and MT MRI metrics change over the course of brain maturation and disentangle what changes in these metrics may indicate on a neurobiological level. Understanding how these MRI metrics change over the course of normal development can potentially yield additional insights into the contrast mechanisms of these metrics and on the underlying mechanisms of changes in disease and injury models.

**Methods**

Detailed information on the data acquisition and analysis pipeline used for this study, including the dataset, is openly available (67). Moreover, the test-retest reproducibility of the MRI metrics have been reported previously (68,69). Therefore, we will only summarize acquisition and analysis and refer the reader to the linked papers for full detailed methods.

*Subjects*

Data used for this study included 12 C57Bl/6 mice (6 males, 6 females) scanned at 3, 4, 5, and 8 months of age. We chose to exclude 'Day 3' and 'Week 1' data from Rahman et al. (67) to avoid possible effects from anesthesia on tissue microstructure. For this study we used data from 11 mice (6 males), as one female mouse did not receive a scan at all timepoints.

*Data Acquisition*

MRI experiments were conducted on a 9.4 Tesla (T) Bruker small animal scanner equipped with a gradient coil insert of 1 T/m strength (slew rate = 4100 T/m/s). During each imaging session, frequency-dependent and tensor-valued dMRI, MT MRI, and anatomical data was acquired with a total scan time of 2 hours and 35 minutes. Anatomical images were acquired using a T2-weighted TurboRARE sequence with parameters: in-plane resolution 150 x 150 µm, slice thickness 500 µm, TE/TR = 40/5000 ms, 16 averages, total scan time of 22 minutes. The frequency-dependent dMRI protocol included a PGSE sequence (gradient duration = 11 ms and diffusion time = 13.8 ms) and OGSE sequences with frequencies of 50, 100, 145, and 190 Hz with a single b-value shell of 800 s/mm$^2$ (10 directions) (70) and parameters: in-plane resolution 175 x 200 µm, slice thickness 500 µm, TE/TR = 39.2/10000 ms, 5 averages, total scan time of 45 minutes. Frequency tuned bipolar (FTB) waveforms were used at 50 Hz to lower the TE of the acquisition (71). The tensor-valued dMRI protocol consisted of LTE and STE acquisitions with b-value shells of 1000 s/mm$^2$ (12 directions) and 2000 s/mm$^2$ (30 directions) with parameters: in-plane resolution 175 x 200 µm, slice thickness 500 µm, TE/TR = 26.8/10000 ms, 3 averages, total scan time of 45 minutes. The MT protocol included three FLASH-3D scans and one B1 map scan to correct for local variations in flip angle. The FLASH-3D scans consisted of an MT-weighted scan, and reference T1-weighted and PD-weighted scans with parameters: in-plane resolution 150 x 150 μm, slice thickness 500 μm, 12 averages, total scan time of 43 minutes.

*Data analysis*

Complex-valued averages were combined using in-house MATLAB code which included frequency and signal drift correction (72) and Marchenko-Pastur denoising of complex-valued data (73). After averages were combined, data underwent correction for Gibbs ringing using Mrtrix3 (74), and eddy-current induced distortions using TOPUP (75) followed by EDDY (76) from FMIRB Software library (FSL, Oxford, UK) (77).

Scalar maps of magnetization transfer ratio (MTR) and MT saturation (Mtsat) were generated from the MT protocol as outlined by Rahman et al. (67). From the frequency-dependent dMRI data, Mrtrix3 was used to fit the diffusion tensor and acquire maps of MD, AD, RD, and FA. Quantitative values of ΔFA, ΔAD, and ΔRD were calculated as the mean of each metric within the region-of-interest (ROI) between the highest frequency (190 Hz) and the lowest frequency (0 Hz). Diffusion dispersion rate (Λ) maps were computed as outlined in Eq. 4. From the tensor-valued dMRI data, maps of $K_{total}$ and $K_{iso}$ were generated by fitting Eq. 1 to the signal from LTE and STE acquisitions, respectively. $K_{aniso}$ and μFA maps were generated using Eq.'s 2 and 3, respectively.

*Region-of-interest (ROI) analysis*

Quantitative MRI parameters were investigated in three regions of interest: global white matter (WM), global deep grey matter (DGM), and the cortex (CX). Masks for these three ROIs were generated from the labelled Turone mouse brain atlas (78). To ensure accurate registration of scalar maps to the atlas, a T2 template, an FA template, and an MT-weighted template was created based on images from all scanning sessions using ANTs software (79). There are 3 steps to warp individual scalar maps to the down sampled atlas space: (1) Individual FA and MT-weighted maps are registered to their respective templates, (2) FA and MT-weighted maps are registered to the T2 template, and (3) the T2 template is registered to the down sampled atlas. Each registration step involves affine transformation followed by symmetric diffeomorphic transformation using ANTs software. Output deformation fields and affine transforms from each of the three steps were used to warp the individual scalar maps to the atlas space to obtain quantitative values for each dMRI metric.

*Statistical Analysis*

To investigate if the MRI metrics changed significantly over time, repeated measures MANOVAs were performed for each metric, over the multiple ROIs (WM, DGM, and CX), to examine differences between timepoints. Metrics with a significant MANOVA were followed up by separate univariate ANOVAs within each ROI, and significant ANOVAs were followed up by Tukey HSD test for post hoc pair-wise comparison to determine differences in metrics within an ROI across time. As previous literature has shown U-shaped trajectories over the healthy brain lifespan (52,60,61,64,80,81), for each ROI and metric, both linear and quadratic models were fit to the data. Furthermore, we performed an extra sum of squares F test to determine whether the data was significantly better fit by a quadratic model as compared to a linear model. Similarly, to examine if the developmental trajectories of male and female mice were significantly different, an extra sum of squares F test was used to determine if both data sets could be accurately fit using a single quadratic fit. Statistical analysis was done in MATLAB and GraphPad Prism version 9.5.1.

**Results**

*MRI Metrics over Time*

**Figure 1** shows representative parameter maps for one mouse at 3 months of age. $K_{aniso}$ shows the same contrast as µFA and enhanced WM contrast compared to $K_{total}$. Λ shows selective enhancement of distinct regions in the brain with densely packed neurons, such as the dentate gyrus (part of the hippocampal formation). MTsat reveals slightly greater contrast than MTR between gray matter and white matter, which is noticeable when comparing the corpus callosum and internal capsule (white matter regions) to the surrounding gray matter.

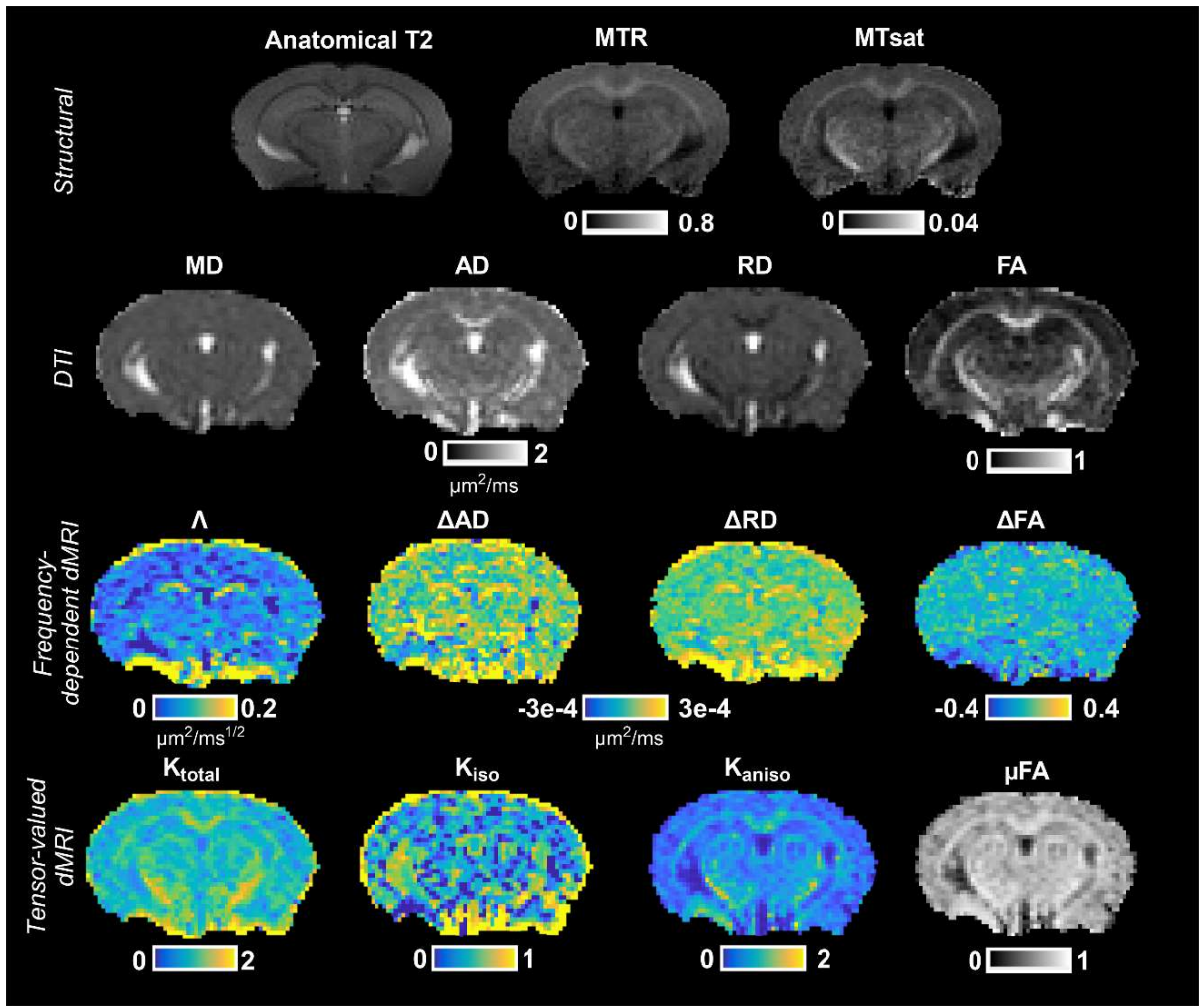

*Figure 1: Representative axial parameter maps from one mouse at 3 months of age.* Structural maps include a T2-weighted map, MTR (magnetization transfer ratio), and MTsat (magnetization transfer saturation). Conventional DTI metrics are shown for reference (MD, AD, RD, and FA). Maps from the frequency-dependent dMRI protocol include Λ (the diffusion dispersion rate), ΔAD, ΔRD, and ΔFA, which show the difference between the DTI metrics at 190 Hz and 0 Hz. Maps from the tensor-valued dMRI protocol include $K_{total}$, $K_{iso}$, $K_{aniso}$, and µFA.

**Figure 2** shows each metric at each timepoint for 3 key ROIs: WM, DGM, and CX. Of the DTI metrics, MD and AD are relatively stable over time, while the significant decreases in RD agree with the significant increases in FA, over all 3 ROIs. Among the frequency-dependent metrics, Λ shows a significant decrease in WM between 4 to 5 months of age. However, ΔAD, ΔRD, and ΔFA do not show any significant changes in follow-up post hoc testing. $K_{total}$ shows significant increases over time for all ROIs, from 3 to 8 months. This is paired with increasing trends of $K_{iso}$, significant in WM and DGM, and increasing trends in MTsat, while $K_{aniso}$ and µFA remain stable over time. Interestingly, while µFA remains stable, FA shows a significantly increasing trend over time.

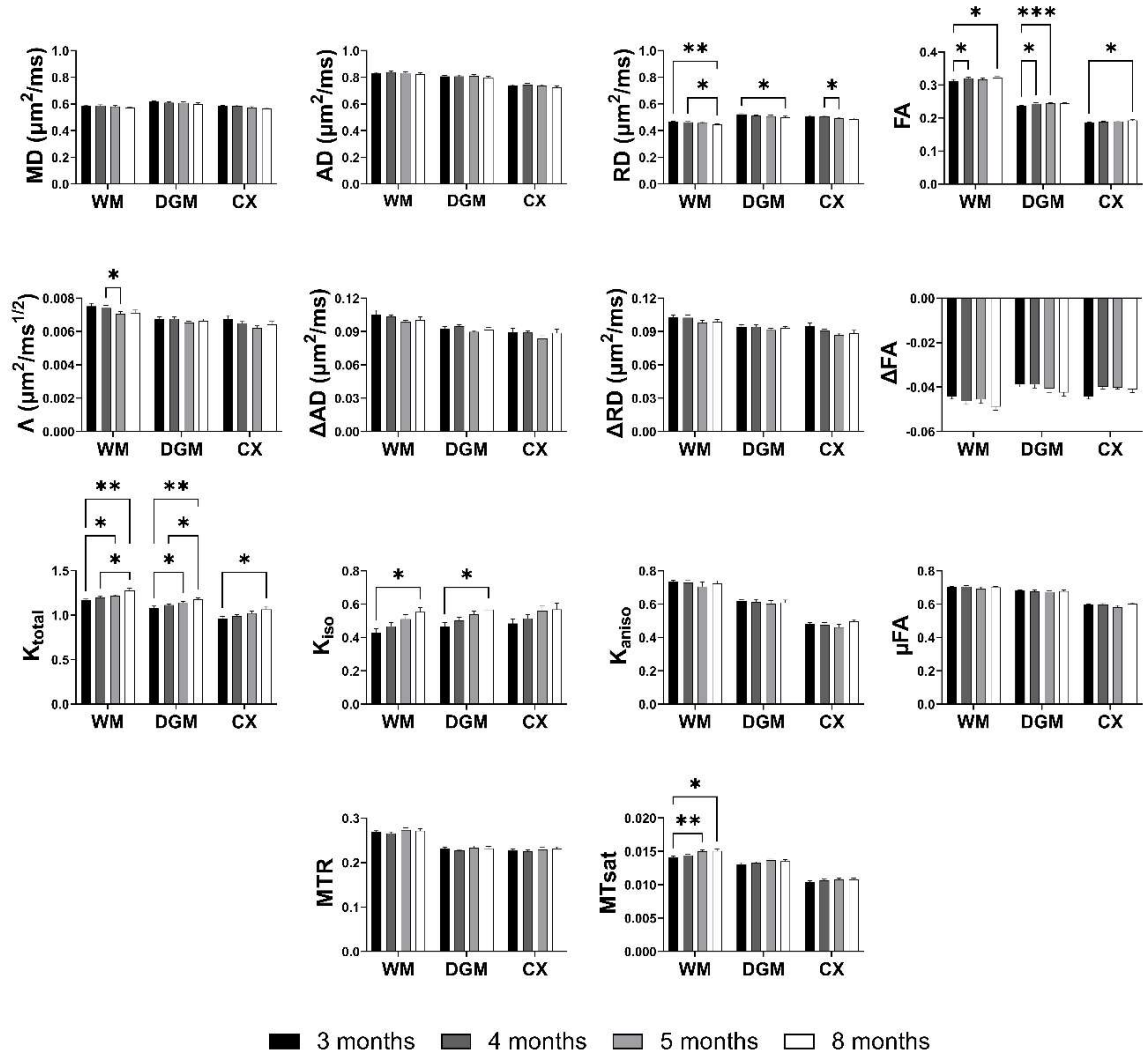

*Figure 2: Quantitative MRI parameter evolution during healthy ageing in white matter (WM), deep gray matter (DGM), and the cortex (CX). Data represents mean values of each metric and error bars represent the standard deviation among n=11 mice. Asterisks represent results from post hoc pair-wise comparison (\* p<0.05, \*\* p<0.01, \*\*\* p<0.001).*

**Supplementary Figure 1** shows each metric at each timepoint for smaller ROIs including 4 white matter ROIs (corpus callosum, internal capsule, external capsule, and fornix) and 4 gray matter ROIs (hippocampus, hypothalamus, thalamus, and amygdala). The trends are similar to the global WM and DGM ROIs, with the corpus callosum and internal capsule showing decreases in RD and increases in MTsat. A trend of increasing $K_{total}$ is found in the corpus callosum, internal capsule, hippocampus, thalamus, and amygdala, while increasing $K_{iso}$ is only observed in the hippocampus, and $K_{aniso}$ and µFA remain stable over time.

*Linear and Quadratic Fits to the Data*

Linear and quadratic fits of all metrics over time, in WM, DGM, and CX are shown in **Figure 3**, with coefficients of determination for each fit and F test results comparing quadratic and linear fits in **Tables 1-3**. Most metrics show a maximum or minimum in the quadratic fits around 5-6 months of age. Significant linear fits ($p < 0.05$) were obtained for MTsat (WM), MD (WM, CX), RD (all ROIs), FA (CX), $K_{total}$ (all ROIs), $K_{iso}$ (WM, DGM), Λ (WM) and ΔFA (WM, DGM). Significant quadratic fits were also obtained for most cases with significant linear fits. In addition to cases with significant linear fits, significant quadratic fits were also obtained for FA (CX and DGM), $K_{iso}$ (all ROIs), Λ (WM and CX), and ΔRD (CX). Although MD did not show any significant changes in **Figure 2**, MD does show a significant linear and quadratic fit over time with a trend of decreasing MD, in WM and CX. For most cases the non-linear fit is not significantly better than the linear fit, except for FA in DGM, ΔRD in cortex, and ΔFA in cortex.

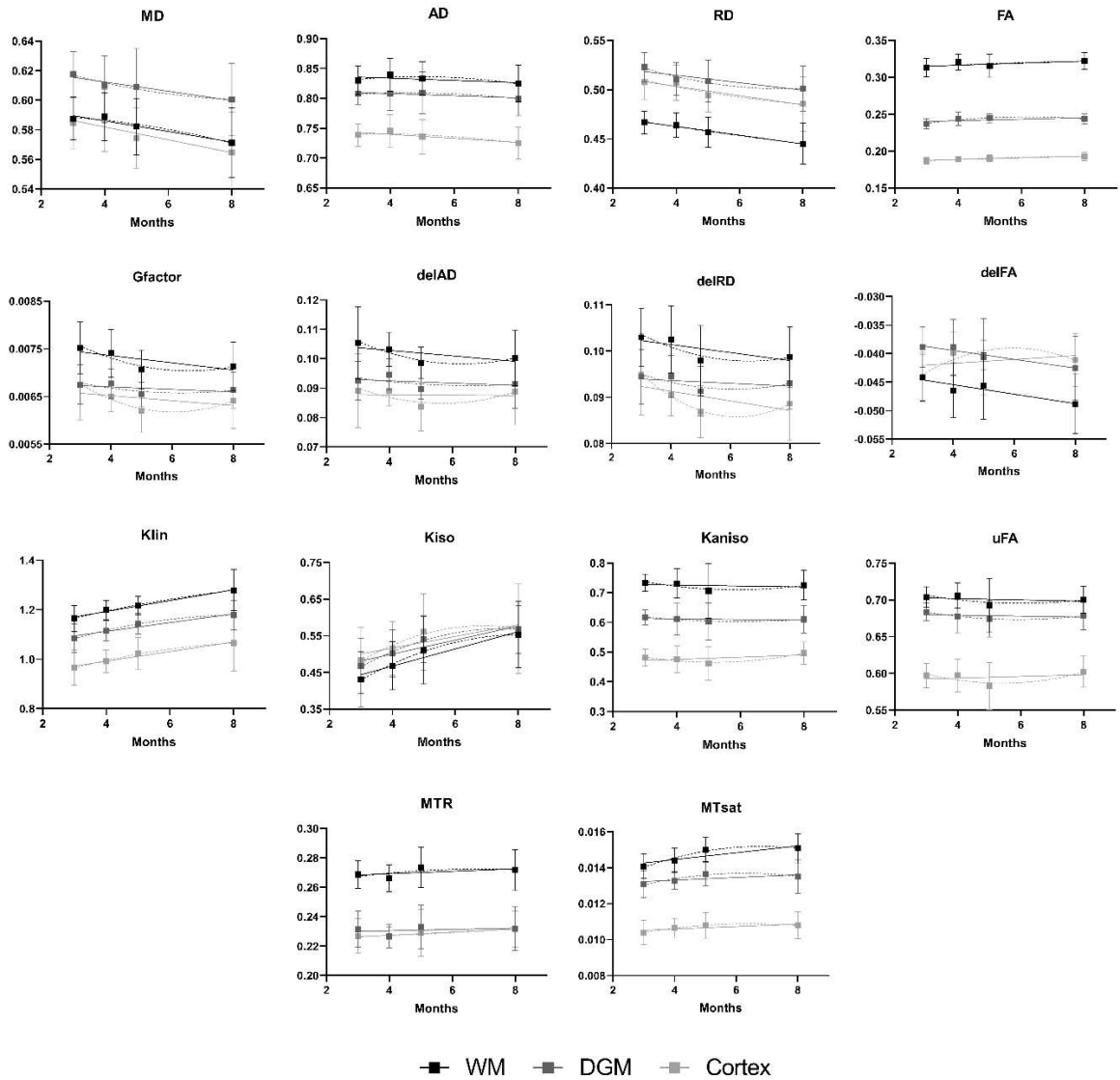

*Figure 3: Linear and quadratic fits of the metrics with age for WM (white matter), DGM (deep gray matter), and CX (cortex).* The solid lines represent linear fits, and the dotted lines represent quadratic fits. For linear and quadratic fits, the coefficient of determination, $R^2$, is reported in Table 1 and Table 2, respectively. Table 3 reports p-values comparing regression models to assess whether the data supports a quadratic model versus a linear model for all metrics in all 3 ROIs.

|  | ROIs | | | | | |
|---|---|---|---|---|---|---|
| Metric | WM | | DGM | | Cortex | |
|  | $R^2$ | P-value | $R^2$ | P-value | $R^2$ | P-value |
| MD | 0.1239 | 0.014078 * | 0.07240 | 0.051452 | 0.1308 | 0.013015 * |
| RD | 0.2386 | 0.00048913 *** | 0.1266 | 0.009886 ** | 0.1674 | 0.004803687 ** |
| AD | 0.01691 | 0.35719545 | 0.01480 | 0.357744 | 0.05840 | 0.104739949 |
| FA | 0.04212 | 0.20224108 | 0.05977 | 0.130435 | 0.1922 | 0.001683248 * |
| Λ | 0.07728 | 0.041764 * | 0.01977 | 0.335896 | 0.03104 | 0.25095 |
| ΔAD | 0.04070 | 0.127422 | 0.01379 | 0.402251 | 0.0002007 | 0.88418 |
| ΔRD | 0.05451 | 0.091653 | 0.01150 | 0.485522 | 0.07185 | 0.093622 |
| ΔFA | 0.09463 | 0.031647 * | 0.07508 | 0.048122 * | 0.02454 | 0.351905 |
| $K_{total}$ | 0.3684 | 1.38E-05 **** | 0.3133 | 2.79E-05 **** | 0.1934 | 0.001705382 ** |
| $K_{iso}$ | 0.2378 | 0.0011176 ** | 0.2279 | 0.00064 *** | 0.08732 | 0.051402541 |
| $K_{aniso}$ | 0.002294 | 0.73969513 | 0.002151 | 0.778384 | 0.02638 | 0.279493007 |
| µFA | 0.005026 | 0.66888684 | 0.003499 | 0.714356 | 0.007399 | 0.533205487 |
| MTR | 0.01870 | 0.427352 | 0.04585 | 0.816788 | 0.02877 | 0.274933 |
| MTsat | 0.1993 | 0.001111 ** | 0.03896 | 0.168421 | 0.03643 | 0.154485 |

*Table 1: Coefficients of determination, $R^2$, and p-values for linear fits to the data over age in WM (white matter), DGM (deep gray matter), and CX (cortex). Highlighted cells show significant fits with p < 0.05.*

|  | ROIs | | | | | |
|---|---|---|---|---|---|---|
| Metric | WM | | DGM | | Cortex | |
|  | $R^2$ | P-value | $R^2$ | P-value | $R^2$ | P-value |
| MD | 0.1263 | 0.047695 * | 0.07531 | 0.149471 | 0.1308 | 0.04733 * |
| RD | 0.2388 | 0.002441 ** | 0.1458 | 0.024483 * | 0.1697 | 0.017217 * |
| AD | 0.0.03110 | 0.461225 | 0.02052 | 0.501572 | 0.06381 | 0.224762 |
| FA | 0.04311 | 0.43848 | 0.1476 | 0.031985 * | 0.2008 | 0.004257 ** |
| Λ | 0.1204 | 0.022125 * | 0.04087 | 0.307299 | 0.1157 | 0.031857 * |
| ΔAD | 0.08250 | 0.051128 | 0.02098 | 0.632942 | 0.03695 | 0.254895 |
| ΔRD | 0.07981 | 0.093223 | 0.03655 | 0.324595 | 0.1696 | 0.009456 ** |
| ΔFA | 0.09464 | 0.097398 | 0.07508 | 0.14487 | 0.1333 | 0.13293 |
| $K_{total}$ | 0.3711 | 8.23E-05 **** | 0.3347 | 7.78E-05 **** | 0.1999 | 0.005298 ** |
| $K_{iso}$ | 0.2552 | 0.001893 ** | 0.2584 | 0.000796 *** | 0.09781 | 0.04666 * |
| $K_{aniso}$ | 0.02334 | 0.274252 | 0.01029 | 0.590635 | 0.07684 | 0.085127 |
| μFA | 0.02271 | 0.276307 | 0.02782 | 0.322506 | 0.06467 | 0.104438 |
| MTR | 0.02473 | 0.510775 | 0.004614 | 0.941344 | 0.02894 | 0.545739 |
| MTsat | 0.2629 | 0.000733 *** | 0.08520 | 0.107656 | 0.05879 | 0.211747 |

*Table 2: Coefficients of determination, $R^2$, and p-values for quadratic fits to the data over age in WM (white matter), DGM (deep gray matter), and CX (cortex). Highlighted cells show significant fits with p < 0.05.*

|         | ROIs   |         |         |
|---------|--------|---------|---------|
| Metric  | WM     | DGM     | Cortex  |
| MD      | 0.7389 | 0.7215  | 0.9993  |
| RD      | 0.9139 | 0.3422  | 0.7392  |
| AD      | 0.4428 | 0.6271  | 0.6288  |
| FA      | 0.8385 | 0.0462* | 0.5108  |
| Λ       | 0.1638 | 0.3478  | 0.0543  |
| ΔAD     | 0.1792 | 0.5863  | 0.281   |
| ΔRD     | 0.2946 | 0.3076  | 0.0337* |
| ΔFA     | 0.9835 | 0.9932  | 0.0287* |
| $K_{total}$ | 0.6796 | 0.2570 | 0.5669 |
| $K_{iso}$   | 0.3335 | 0.2014 | 0.2744 |
| $K_{aniso}$ | 0.3528 | 0.5647 | 0.1420 |
| µFA     | 0.3940 | 0.3171  | 0.1208  |
| MTR     | 0.6173 | 0.9729  | 0.9330  |
| MTsat   | 0.0670 | 0.1576  | 0.3295  |

*Table 3: P-values from extra sum of squares F test for each metric in WM, DGM, and Cortex showing if the non-linear fit is significantly better than linear fit.* Highlighted cells show significantly better non-linear fits with $p < 0.05$

*Sex-dependent Differences over Time*

Data was separated by sex to examine sex-based differences for all metrics in WM, CX, and DGM (**Figure 4**). Most metrics show significantly different fits between males and females. For the diffusivity metrics (MD, AD, RD), the females show an inverted U-shape trajectory, while the males consistently show a decreasing trend. Among the frequency-dependent metrics, ΔAD and ΔFA show the same fits for both sexes, while Λ shows a U-shape trajectory for females and an inverted U-shape for males in the CX, driven by the ΔRD trajectories. $K_{total}$ and $K_{iso}$ both show a U-shape trajectory for males and an inverted U-shape trajectory for females. Interestingly, $K_{iso}$ and $K_{aniso}$ show opposing trends for males and females. As expected, $K_{aniso}$ and µFA show similar trends to each other for both sexes. MTR and MTsat also show the same fits for both sexes, with MTsat showing an increasing trend up to 5 months and remaining stable until 8 months.

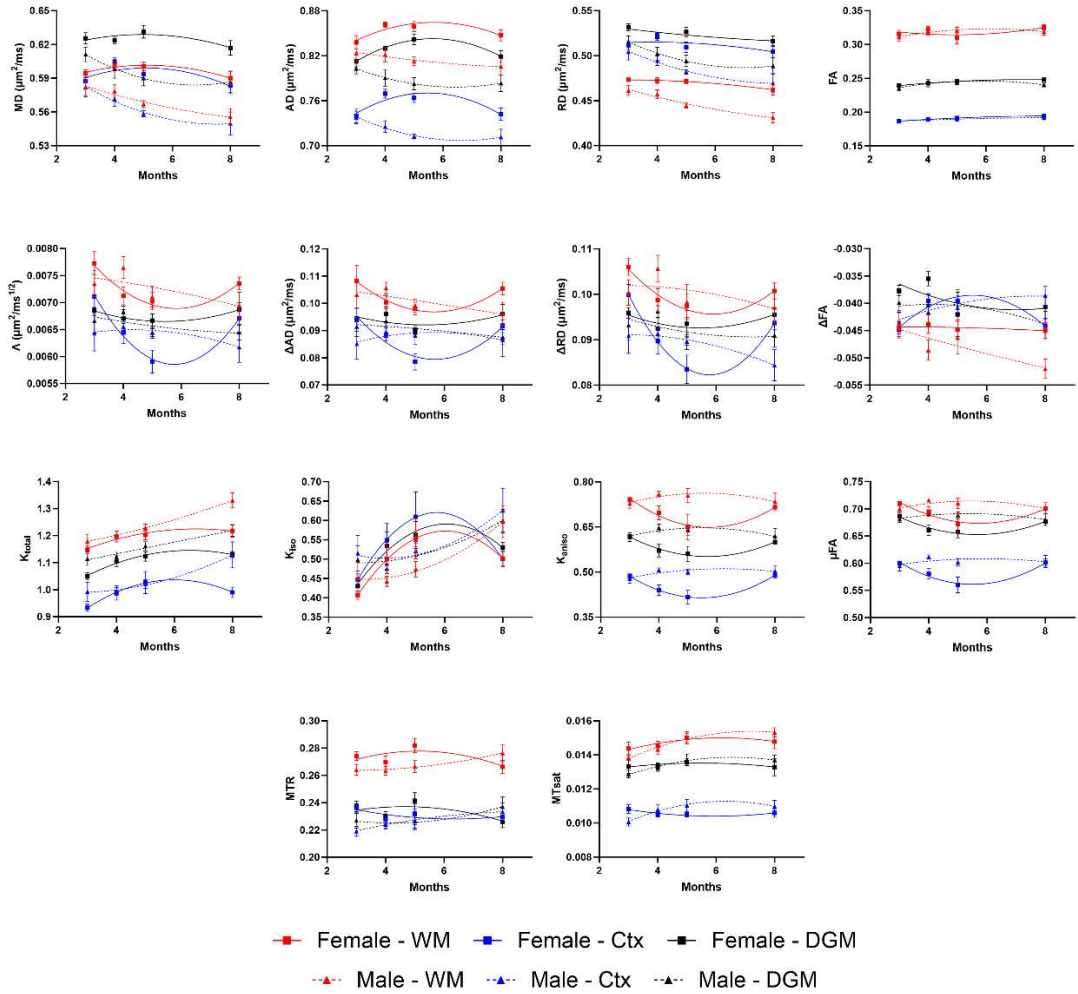

*Figure 4: Plots showing quadratic fits to the data, separated by sex, for each metric in WM (white matter), DGM (deep gray matter), and CX (cortex).* For those plots with separate fits for male and female, there was a significant difference in the fitting parameters, and for those with a single line, a single fit could accurately represent both datasets.

*Linear Regression of Kurtosis with Myelin-Specific Metrics*

To explore how kurtosis relates to the myelin-specific metrics over time, **Figure 5** shows linear regressions of the kurtosis metrics with MTR and MTsat over all timepoints and mice, for each ROI. $K_{total}$ and $K_{iso}$ show similar trends, with positive correlations with MTR and MTsat in the WM and CX. Although $K_{aniso}$ does not show any correlations with MTsat, negative correlations are found for $K_{aniso}$ and MTR for all ROIs.

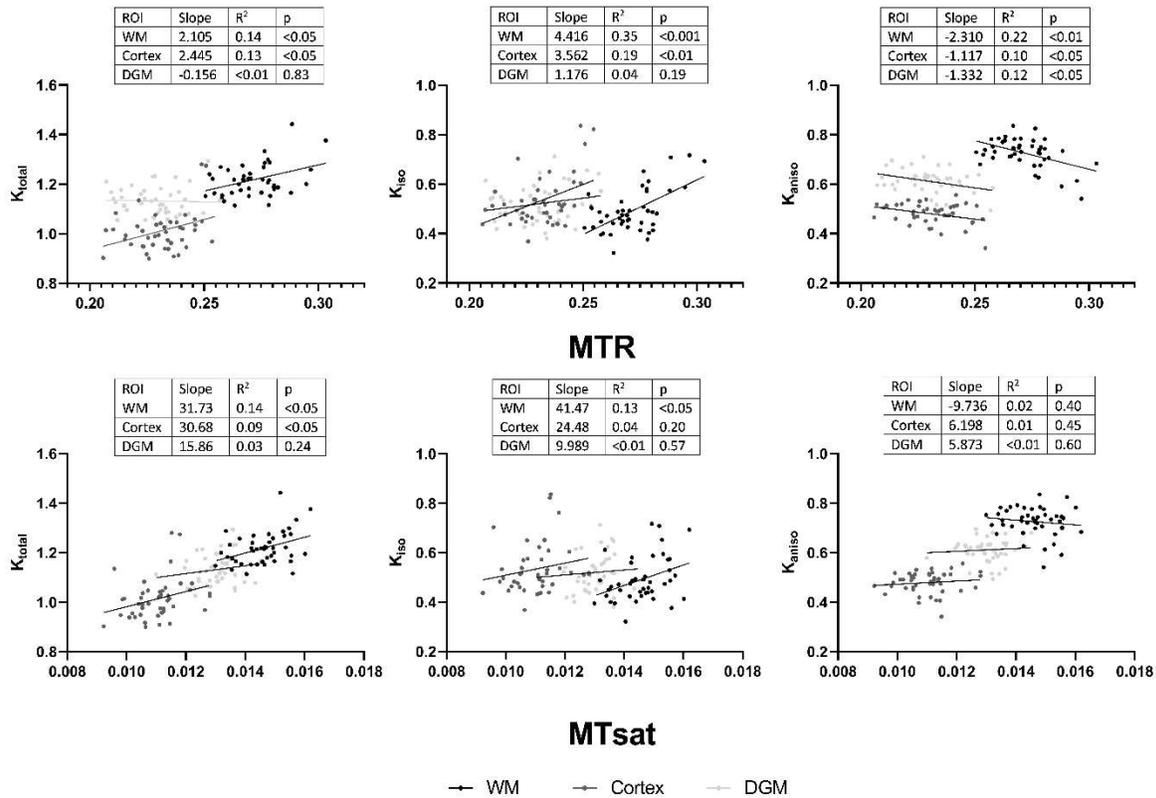

*Figure 5: Plots illustrating linear fits of the kurtosis metrics ($K_{total}$, $K_{iso}$, and $K_{aniso}$) to the myelin-specific metrics (MTR and MTsat) in WM (white matter), DGM (deep gray matter), and CX (cortex) over all timepoints.* The slope, $R^2$, and p-value are reported for each fit.

To explore the linear relationship of kurtosis metrics to myelin-specific metrics from gray to white matter, **Figure 6** shows the linear fits of the kurtosis metrics with MTR and MTsat over all ROIs. Interestingly, $K_{aniso}$ shows a positive correlation with both MTR and MTsat, in contrast to **Figure**

**5**. The positive correlation of $K_{total}$ with MTR and MTsat is driven by the trend in $K_{aniso}$, as $K_{iso}$ does not show any correlation with MTR and MTsat over all ROIs.

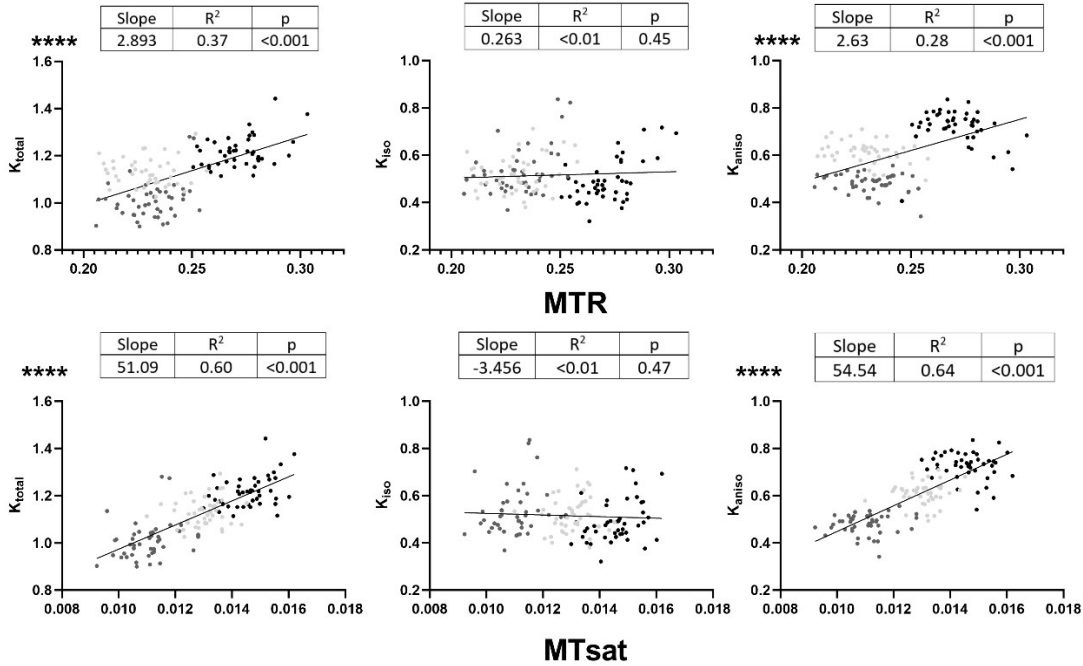

*Figure 6: Plots illustrating linear fits of the kurtosis metrics ($K_{total}$, $K_{iso}$, and $K_{aniso}$) to the myelin-specific metrics (MTR and MTsat) for all ROIs together (WM, DGM, and Cortex) over all timepoints.* The slope, $R^2$, and p-value are reported for each fit.

**Discussion**

Here we explored healthy mouse brain maturation using frequency-dependent and tensor-valued dMRI and MT MRI to probe smaller spatial scales, orientation-independent diffusion and various kurtosis sources, and myelin content, respectively. Our protocols were applied longitudinally in mice between 3 – 8 months of age to better understand the contrast mechanisms of these advanced MRI metrics and how they evolve over the course of healthy brain maturation. As microstructural MRI is becoming more prevalent in the neuroscience community, there is a need to assess the evolution of these metrics in healthy rodents, which may provide insights into the underlying mechanisms of changes in these metrics in disease and injury models.

Although there have been several human lifespan imaging studies, healthy rodent brain maturation studies are sparse. This study builds on previous rodent studies (5,6,65,66), including more timepoints and a more comprehensive set of advanced imaging protocols to probe the evolving microstructure in the maturing brain. Importantly, this is the first study to characterize the evolution of frequency-dependent and tensor-valued dMRI metrics longitudinally during brain maturation. Additionally, we show here changes in total kurtosis over time may be dominated by isotropic sources, which suggests that total kurtosis is unlikely to be sensitive to how myelination affects water diffusion anisotropy over the course of normal brain development but may rather be sensitive to myelin formation/remodelling and glial cell diversification. For studies that only calculate total kurtosis (likely as most only use LTE acquisitions), we suggest caution in attributing neurobiological changes to changes in total kurtosis as we show here no changes in anisotropic sources of kurtosis in the presence of increasing myelin content.

*MRI Metrics over Time*

The trends observed in our DTI metrics are comparable to previous human and rodent studies investigating normal brain development (5,6,52,63,64,80,81). As MD and AD do not show any significant change with post hoc analyses (although they show significant time effects and a decreasing trend with age), the significant increases in FA in WM, DGM, and CX are driven by the significant decreases in RD in these three regions between 3 – 8 months (**Figure 2**). Hammelrath et al. reported that in most white matter ROIs, FA continuously increased in agreement with increased intensity on myelin staining, in mice studied until 6 months of age (5). Mengler et al. found a pronounced increase in myelinated fibers, using histology, in the cortex between 3 – 6 months of age in rats (6). They interpreted that the drop in MD and RD with age was likely due to myelination. Myelination results in reduced RD, as radial water diffusivity becomes more restricted, which consequently leads to reduced MD (82). In addition, myelination, as well as neurogenesis (83), decreases the extracellular free water compartment in favour of the intracellular compartment, which can also lower MD (84). Interestingly, in a histological study, Mortera et al. found a significant trend of widespread and progressive neuronal loss that began as early as 3 months of age in rats (15-20 % decrease in neuron counts between 3 – 5 months of age in the cortex and olfactory bulb), when neuronal numbers are maximal in all structures, which indicates that age-related decline in the brain begins as soon as the end of adolescence (83). Neuronal loss would result in increased MD, and counteract the decreased MD due to myelination, which may explain why no significant changes are observed in MD after post hoc analyses (**Figure 2**). Additionally, Mortera et al. reported that brain mass increased in all structures over time despite a decline in numbers of neurons, which implied that the average size of the surviving cells, neuronal and/or non- neuronal, increased during aging.

Among the metrics from the frequency-dependent dMRI protocol, there is a significant decrease in Λ, the diffusion dispersion rate, between 4 – 5 months of age in WM (**Figure 2**). As Λ shows similar contrast to ΔMD (68), we can refer to previous ΔMD studies to interpret changes in Λ. Aggarwal et al. reported elevated ΔMD contrast in mouse brain regions with high neuron packing, such as in the dentate gyrus of the hippocampus (85). This elevated contrast in the dentate gyrus can also be seen in the Λ map in **Figure 1**. In a phantom study, Parsons et al. related elevated ΔMD to the presence of larger spherical structures (86), and in a simulation study, Xu et al. related elevated ΔMD with higher nuclear volume fraction (87). In another study, Aggarwal et al. found consistent decrease in ΔMD in the CA1 Pyramidal Layer of the hippocampus in healthy mice during the first two months after birth and related this to a progressive increase in pyramidal cell soma size with age (88). Thus, the decrease in Λ observed in **Figure 2** can be interpreted as a reduction in axonal packing, which is also supported by the neuron loss finding of Mortera et al. (83) and may be related to synaptic pruning during development (89), and an increase in soma cell size. The other frequency-dependent metrics here do not show any significant changes after post hoc analyses, possibly due to higher inter-subject variation in these metrics, although ΔRD did show a significant time effect, with decreasing trends in WM and CX. Aggarwal et al. found significantly higher ΔRD in the mouse corpus callosum in a cuprizone demyelination model (85), so the decreasing trend of ΔRD here may be due to continued myelination.

$K_{total}$ shows significant increases over time for WM, DGM, and CX, from 3 to 8 months. This is comparable to previous DKI studies in healthy brain maturation (56,65,66), and previous literature has related $K_{total}$ increases to myelination and dense packing of axon fibers in white matter and more densely packed structures like cells or membranes and dendritic architectural modifications in gray matter. However, these interpretations involving fibre microstructure would be expected to coincide with changes in anisotropic components of kurtosis, which was not seen here. By investigating both isotropic and anisotropic kurtosis components in this study, we observe that during healthy brain development, changes in $K_{total}$ are driven by changes in $K_{iso}$ (significant in WM and DGM). This is paired with increasing trends in MTsat, which indicates increasing myelination, while $K_{aniso}$ and µFA remain relatively stable over time. MTR does not show any significant trends over time, which could be due to the reduced myelin-sensitivity of MTR, while MTsat has improved specificity to myelin (47). Moreover, the contrast between gray and white matter regions, averaged over all mice in this dataset, was previously shown to be significantly higher in MTsat, compared to MTR (69). As there is a trend of increasing $K_{total}$, $K_{iso}$, and MTsat, the relationship of the kurtosis metrics and the myelin-specific metrics is further explored in **Figure 5** and **Figure 6**.

The mean µFA values are much higher than mean FA values (by ~0.4), indicating there is orientation dispersion of axons within our ROIs. It is interesting to note that although µFA remains stable, FA is increasing over time, which may be due to fibers becoming more aligned through development. As fibers become more aligned, FA is expected to increase, as macroscopic water diffusion anisotropy increases, but µFA is not expected to change, as it disentangles effects of fiber orientation dispersion from microstructure (20,25). As µFA remains unchanged over time, this indicates that axon integrity is not compromised over this period. Moreover, as there are no changes in µFA despite increases in MTsat, this suggests myelination in the mature brain is not a main contributor to microscopic diffusion anisotropy and anisotropic kurtosis in axons. In a study investigating axon diameters and myelin content in fixed rat spinal cord, a moderate negative correlation was observed between µFA and MWF (myelin water fraction), while FA was positively correlated with MWF (51). This is comparable to our study as increasing FA is paired with increasing MTsat. Shemesh explains that the negative correlation between µFA and MWF would reflect indirectly the approximately constant g-ratio in healthy tissue, rather than enhanced restriction, and as axon diameter increases with myelination, µFA decreases (51). Additionally, unlike FA, compartments of proportionally different sizes can give rise to the same µFA. However, µFA reported in the rat spinal cord study was derived from Double Oscillating Diffusion Encoding (DODE) protocols, which can probe smaller spatial scales. Thus, the µFA calculated from our b-tensor encoding protocol may not be sensitive to changes in axon diameter, as the larger the diffusion time, the longer path will be probed in the unrestricted dimension (parallel to the axon length), and the µFA will be larger and less reflective of axon diameter, and consequently, myelination.

The increase in $K_{iso}$ reflects increased cell size and density heterogeneity, as it indicates a heterogeneous microenvironment with compartments having a wide variation of diffusivities within the voxel (90,91). We hypothesize that the increased cell size/density heterogeneity may be due to (1) differentiation and maturation of oligodendrocytes (OLs) and/or (2) increased number and increased diversity of glial cells. Oligodendrocyte precursor cells (OPCs) divide and generate myelinating oligodendrocytes (OLs) throughout postnatal and adult life. Although most OPC proliferation, OL differentiation, and myelin development occurs before 3 months of age, after postnatal 3 months (the age of our mice at the start of this study), OPCs continue to proliferate, OLs continue to differentiate, and adaptive myelination continues, at a lower stable rate (92). He et al. showed increased $K_{iso}$ in LPC (L-α-Lysophosphatidylcholine) treated regions in the mouse brain (93). LPC is known to kill mature OLs, so increased $K_{iso}$ could be due to re-population of mature OLs by OPCs, which has been seen 3-7 days post-LPC treatment (94). The increase in $K_{iso}$ may reflect oligodendrocyte proliferation and maturation, and thus myelin formation, which is supported by the increase observed in MTsat. Elevated $K_{iso}$ may also be related to the increasing release of myelin debris over the course of normal aging, which is effectively cleared by activated microglia (95), also known to play a role in 'myelin remodeling'. $K_{iso}$ may be sensitive to microglia activation associated with normal brain development, which would result in diversification of cell morphology. Cell size and density of astrocytes increasing at different rates in different regions of the brain may also result in increased $K_{iso}$ (96,97). In the rat cerebral cortex, Sabbatini et al. reported that the size and number of astrocytes increased progressively from 3 to 24 months old (97). Age-associated increases in astrocytic hypertrophy and microglial activation were found to be prominent in white matter (96), which may explain why we observe the largest increase of $K_{iso}$ in WM. It is likely that changes in $K_{iso}$ are reflective of these various microstructural changes to some degree, as increases in $K_{iso}$ in GM may suggest glial cell diversification whereas the larger increase in WM may suggest a combination of glial cell diversification including OL proliferation/maturation.

*Linear and Quadratic Fits to the Data*

Quadratic fits of the metrics with age (**Figure 3**) reveal a maximum or minimum for most metrics around 5-6 months of age. This is comparable with the mean kurtosis peak of 6 months in rats reported by Han et al. (66). Although most metrics that show a significant linear fit (**Table 1**) also show a significant quadratic fit (**Table 2**), the quadratic fit is not significantly better than the linear fit in most cases (**Table 3**). Investigating more subjects and timepoints after the minimum/maximum had been reached would likely have improved the quadratic fits. The largest variability is also observed around the maximum/minimum period, likely due to individual mice reaching their maxima/minima at slightly different times.(52,54,60,61,64) Among the frequency-dependent metrics, significant quadratic fits were found for Λ and ΔRD. Interestingly, a significant linear trend of decreasing ΔFA is found in WM and DGM. To our knowledge, previous literature has not discussed values of ΔFA, and our results show that ΔFA may be a relevant frequency-dependent metric to explore. Significant linear and quadratic fits were found for $K_{total}$ and $K_{iso}$, but not for $K_{aniso}$, which indicates that $K_{aniso}$ is not sensitive to age-related changes during normal brain maturation.

The linear and quadratic fits shown here are comparable to previous studies in both humans (52,54–56,60,61,63,64,81) and rodents (66), investigating DTI, DKI, and myelin-specific metrics. In a human lifespan study, peak age calculated with the quadratic model revealed peak ages in the range of 30-50 years depending on brain region and metric (64). 30 human years is equivalent to about 6 mouse months and 50 human years is equivalent to about 15 mouse months (98). This is consistent with our results, as we start to observe peaks around the 5-6 month timepoint. Most DTI studies indicate a U-shaped relationship between the diffusivity metrics and age in most ROIs, with FA exhibiting an inverted U-shaped relationship. An inverted U-shape trend of myelin-specific metrics with age has been reported (60,61,64,99–101), and this quadratic association is attributed to the process of myelination from youth through middle age, followed by demyelination in later years.

*Sex-dependent Differences over Time*

Sex-dependent differences over time are observed in most of the dMRI metrics (**Figure 4**), while MTsat shows the same fits for both sexes. This suggests that the diffusion metrics are more sensitive to differences between sexes, as they are sensitive to other factors beyond myelination. Interestingly, the overall shape of the fits for the metrics do not change between ROIs (WM, CX, DGM), possibly as the ROIs are quite large. MD, AD, and RD show opposing trends for both sexes, with males following a U-shape trajectory. Among human lifespan studies, Kiely et al. (64) reported that sex effects were limited to very few ROIs, while Grydeland et al. (60) and Lebel et al. (52,81) found no sex effects. Although these studies included both sexes, some studies contained disproportionately more males, while others contained almost equal proportions, but it is unknown what percentage of males and females were considered for each age range, which may have introduced bias into the results. According to a review, most DTI human lifespan studies have not explicitly measured sex differences over time (80). Previous DTI studies (imaged at a single cross-sectional timepoint) have found higher FA and lower MD in males (the lower MD is consistent with our results), tied to hormonal levels, although contradictory and null findings have also been reported (102). The frequency-dependent metrics show the same fit for males and females for most cases in **Figure 4**. Tetreault et al. were the first to explore sex differences in the human corpus callosum using frequency-dependent dMRI (103). They reported greater ΔRD in the genu of males, which could reflect larger axon diameters than females. However, this study combined data from all ages (20-73 years old) when reporting this difference, so analyzing age ranges separately may change the results.

Kurtosis metrics and µFA also show opposing trends for both sexes, with females showing an inverted U-shape for $K_{total}$ and $K_{iso}$ and a U-shape trajectory for $K_{aniso}$ and µFA. This further supports our finding that $K_{total}$ changes over time are driven by changes in $K_{iso}$. Previous µFA studies have not explored sex differences, as most studies have been proof-of-concept. The differences in $K_{iso}$ trajectories between sexes may be related to differences in glial cell number and morphology and/or oligodendrocyte cell diversity between males and females. Mouton et al. reported that female mice have significantly higher numbers of microglia and astrocytes than males (consistently as measured from 3-24 months) in the hippocampus (104). Previous studies have found that female microglia are more developmentally mature than male microglia and females have higher expression of inflammatory, phagocytic, and immune genes than males (105). Thus, females could have lower levels of cell proliferation due to increased phagocytosis of progenitor cells, which may explain the overall lower trend of $K_{total}$ and $K_{iso}$ in the female trajectory. The differences seen in the male and female trajectories in **Figure 4** highlight the importance of including both sexes in research and considering sex-dependent analyses. However, these differences may be amplified by the oldest age lying only barely outside the peak of the U-shape, and may be consistent with a later peak in males (that is not captured here) compared to females. Combined with the small sample size in this study, it is challenging to fully interpret the sex dependent trajectories, which will require larger sample sizes and age ranges, and further histological analyses.

*Linear Regression of Kurtosis with Myelin-Specific Metrics*

**Figure 5** illustrates the linear relationship of kurtosis metrics with myelin-specific metrics in each ROI independently. $K_{aniso}$ does not show any correlations with MTsat (**Figure 5**). The negative correlations of $K_{aniso}$ with MTR may include biases, as MTR is sensitive to T1-weighting and various sequence parameters. This indicates that $K_{aniso}$ (and µFA via Eq. 3) is not sensitive to myelination, which is in agreement with the results in **Figure 2**. Although previous studies found changes in µFA with demyelination in the cuprizone demyelination model (93) and multiple sclerosis patients (27), we anticipate that subsequent damage to axons and/or neuroinflammation may have caused resultant changes in µFA and $K_{iso}$ as well, which supports our reasoning that myelination may not be a key contributor to axonal anisotropy. **Figure 5** shows that increases in $K_{total}$ over the course of development, which have been shown by others in humans (55,56) and rodents (65,66), are largely due to increases in isotropic kurtosis rather than anisotropic kurtosis. Furthermore, the results highlight the importance of using both LTE and STE acquisitions as $K_{total}$ accounts for kurtosis from both anisotropic and isotropic sources, and most studies only explore the total kurtosis. We show here changes in total kurtosis over time may be predominated by isotropic sources, which suggests that total kurtosis is unlikely to be sensitive to how myelination affects water diffusion anisotropy over the course of normal brain development (as these changes would be reflected in anisotropic sources of kurtosis).

**Figure 6** illustrates the linear fits of kurtosis metrics with myelin-specific metrics in all ROIs combined, to focus on linear relationships over the ROIs. As $K_{aniso}$ shows positive correlations with both MTR and MTsat (**Figure 6**), and $K_{iso}$ does not show any correlations here, this indicates that $K_{aniso}$ is sensitive to microstructural differences that distinguish different brain regions, while $K_{iso}$ is not. **Figure 5 and Figure 6 highlight that changes in total kurtosis found over time (in normal brain development) are driven by isotropic kurtosis, while differences in total kurtosis found between brain regions are driven by anisotropic kurtosis.** This indicates that while the main differences between white and gray matter regions stem from fiber content and alignment, leading to differences in anisotropic kurtosis, the main differences over time are not related to changes in axonal content. Age and region dependent kurtosis changes have been reported in previous DKI studies (55,56,65,66,106–108), but these studies only focused on the total kurtosis.

*Limitations*

Although conducted using a longitudinal study design and state-of-the-art methods, our investigation has limitations. Firstly, we do not calculate microscopic kurtosis (μK), which is another source of total kurtosis and ignoring it can impact the accuracy of other kurtosis sources. μK is the weighted sum of different microscopic sources of non-Gaussian diffusion, which include restricted diffusion inside compartments, microstructural disorder due to the presence of microscopic hindrances to water molecules (such as membranes and axon caliber variations), and exchange between components (109,110). Recently, μK was shown to be a primary driver of total kurtosis upon ischemia in mice (111) and was mapped in human brain tissue for the first time (112), revealing that this component is non-negligible. Moreover, Novello et al. showed that assuming the multiple Gaussian component approximation for kurtosis source estimation (ignoring μK) introduces significant bias in the estimation of other kurtosis sources (112). Additionally, for our frequency-dependent dMRI protocol, our highest OGSE frequency (190 Hz) was determined by hardware constraints, and reaching even higher frequencies would allow us to probe smaller spatial scales. Although our MTsat protocol improves myelin specificity compared to MTR, a more recent technique, inhomogeneous magnetization transfer (ihMT), may be better suited for our study. ihMT, which was developed based on MT MRI, is more specific to myelin than MTsat, due to its direct sensitivity to the phospholipids in myelin (113,114). In this study, we only explored one phase of the mouse brain lifespan, the brain maturation phase, while the degeneration phase remains to be investigated. Including more timepoints after 8 months would allow for a more robust and complete picture of the mouse brain lifespan trajectory. However, the time period used in this study (between 3-8 months of age) is a widely used time period for longitudinal rodent neuroimaging studies and will provide insight into healthy rodent brain development to help disentangle normal and pathological microstructural changes. It is also important to note that the developmental trajectory described here is for wild type C57Bl/6 mice and may be altered in other mouse lines.

**Conclusions**

In conclusion, we investigated the evolution of advanced dMRI and MT MRI metrics longitudinally in healthy mouse brain maturation, as the study of normal brain maturation will help exclude confounds of cerebral developmental changes from interpretations of disease and injury mechanisms. Overall, the trends observed in our DTI and MTsat metrics are comparable to previous human and rodent studies investigating normal brain development. Neurobiological changes that result in changes to isotropic kurtosis remain understudied, however, we suggest here that isotropic kurtosis sources drive changes in total kurtosis during normal brain maturation. Our results suggest myelination is not a main contributor to microscopic diffusion anisotropy and anisotropic kurtosis in axons. For studies that only calculate total kurtosis, we suggest caution in attributing neurobiological changes to changes in total kurtosis as we show here no changes in anisotropic sources of kurtosis in the presence of myelination.

**Author contributions**

**Naila Rahman:** Conceptualization, Data curation, Formal analysis, Investigation, Visualization, Writing – original draft, Writing – review & editing. **Jake Hamilton:** Conceptualization, Formal analysis, Investigation, Visualization, Writing – original draft, Writing – review & editing. **Kathy Xu:** Project administration, Resources, Writing – review & editing. **Arthur Brown:** Funding acquisition, Project administration, Resources, Supervision, Writing – review & editing. **Corey A. Baron:** Conceptualization, Funding acquisition, Resources, Supervision, Writing – review & editing.

**Declaration of competing interests**

The authors declare no competing interests.